\documentclass[aps,prb,reprint,superscriptaddress,longbibliography,amsmath,amssymb]{revtex4-1}

\usepackage{graphicx}
\usepackage{dcolumn}
\usepackage{bm}
\usepackage{soul}
\usepackage{color}
\usepackage{float}

\usepackage{enumitem}
\usepackage{setspace}

\usepackage{helvet}

\renewcommand{\vec}{\bf}

\begin{document}

\title{De novo exploration and self-guided learning of potential-energy surfaces}

\author{Noam Bernstein}
\affiliation{Center for Materials Physics and Technology, 
U.S. Naval Research Laboratory, Washington, DC 20375, United States}
 
\author{G\'{a}bor Cs\'{a}nyi}
\affiliation{Department of Engineering, University of Cambridge, 
Cambridge CB2 1PZ, United Kingdom}

\author{Volker L. Deringer}
\email{vld24@cam.ac.uk}
\affiliation{Department of Engineering, University of Cambridge, 
Cambridge CB2 1PZ, United Kingdom}
\affiliation{Department of Chemistry, University of Cambridge,
Cambridge CB2 1EW, United Kingdom} 

\date{\today}

\renewcommand{\abstractname}{\vspace{5mm}}

\begin{abstract}
Interatomic potential models based on machine learning (ML) are rapidly developing as tools for materials simulations. 
However, because of their flexibility, they  require large fitting databases that are normally created with substantial manual selection and tuning of reference configurations. 
Here, we show that ML potentials can be built in a largely automated fashion, exploring and fitting potential-energy surfaces from the beginning ({\em de novo}) within one and the same protocol. 
The key enabling step is the use of a configuration-averaged kernel metric that allows one to select the few most relevant structures at each step.
The resulting potentials are accurate and robust for the wide range of configurations that occur during structure searching, despite only requiring a relatively small number of single-point DFT calculations on small unit cells.
We apply the method to materials with diverse chemical nature and coordination environments, marking a milestone toward the more routine application of ML potentials in physics, chemistry, and materials science.
\end{abstract}

\maketitle

\section*{Introduction}

Atomic-scale modeling has become a cornerstone of scientific research.
Quantum-mechanical methods, most prominently based on density-functional theory (DFT), describe the atomistic structures and physical properties of materials with high confidence \cite{Lejaeghere2016};
increasingly, they also make it possible to discover previously unknown crystal structures and synthesis targets \cite{Oganov2019}. 
Still, 
quantum-mechanical materials simulations are severely limited by their high computational cost.

Machine learning (ML) has emerged as a promising approach to tackle this long-standing problem \cite{Behler2007, Bartok2010, Li2015, Artrith2016, Smith2017, Chmiela2017, Behler2017, Huan2017, Chmiela2018, Zhang2018a}. ML-based interatomic potentials approximate the high-dimensional potential-energy surface (PES) by fitting to a reference database, which is usually computed at the DFT level. 
Once generated, ML potentials enable accurate simulations that are orders of magnitude faster than the reference method.
They can solve challenging structural problems, as has been demonstrated for the atomic-scale deposition and growth of amorphous carbon films \cite{Caro2018}, for proton-transfer mechanisms \cite{Hellstrom2019} or dislocations in materials \cite{Fellinger2018, Maresca2018}, involving thousands of atoms in the simulation. 
More recently, it was shown that ML potentials can be suitable tools for global structure searches targeting  crystalline phases \cite{C_AIRSS, B_GAP, GAP-RSS_Faraday, Podryabinkin2019}, clusters \cite{Ouyang2015, Tong2018, Kolsbjerg2018, Hajinazar2019}, and nanostructures \cite{Eivari2017}. 

Assembling the reference databases to which ML potentials are fitted is currently mostly a manual and laborious process, guided by the physical problem under study. 
For example, hierarchical databases for transition metals have been built that start with simple unit cells and gradually add relevant defect models \cite{Szlachta2014, Dragoni2018}; liquid and amorphous materials can be described by iteratively grown databases that contain relatively small-sized MD snapshots \cite{Eshet2010, Sosso2012, amoC, Mocanu2018}. 
A ``general-purpose'' Gaussian Approximation Potential (GAP) ML model for elemental silicon was recently developed \cite{Bartok2018} which can describe crystalline phases with meV-per-atom accuracy, treat defects, cracks, and surfaces \cite{Bartok2017}, and generate amorphous silicon structures in excellent agreement with experiment \cite{aSi_structures_GAP}. 
Despite their success in achieving their stated goals, none of these potentials are expected to be even reasonable for crystal 
structures not included in their databases, say, phases that are stable at very high pressures. 

In contrast, structure searching (that is, a global exploration of the PES) can be a suitable approach for finding structures to be included in the training databases {\em in the first place} \cite{Hajinazar2017, B_GAP, GAP-RSS_Faraday, Podryabinkin2019}. 
The principal idea to explore configuration space with preliminary ML potentials is well established: since the first high-dimensional ML potentials have been made, it was shown how they can be refined by exploring unknown structures \cite{Behler2007, Behler2008, Sosso2012}, and ``on the fly'' schemes were proposed to add required data while an MD simulation is being run \cite{Li2015, Junnouchi2019, Vandermause2019, Junnouchi2019a}.
 We have previously shown that the PES of boron can be iteratively sampled without prior knowledge of any crystal structures involved; we called the method ``GAP-driven random structure searching'' (GAP-RSS) \cite{B_GAP}, reminiscent of the successful Ab Initio Random Structure Searching (AIRSS) approach \cite{Pickard2006, Pickard2011}. Subsequently we demonstrated, by way of an example, that the crystal structure of black phosphorus can be discovered by GAP-RSS within a few iterations, and we identified several previously unknown hypothetical allotropes of phosphorus \cite{GAP-RSS_Faraday}.

\begin{figure*}
    \centering
    \includegraphics{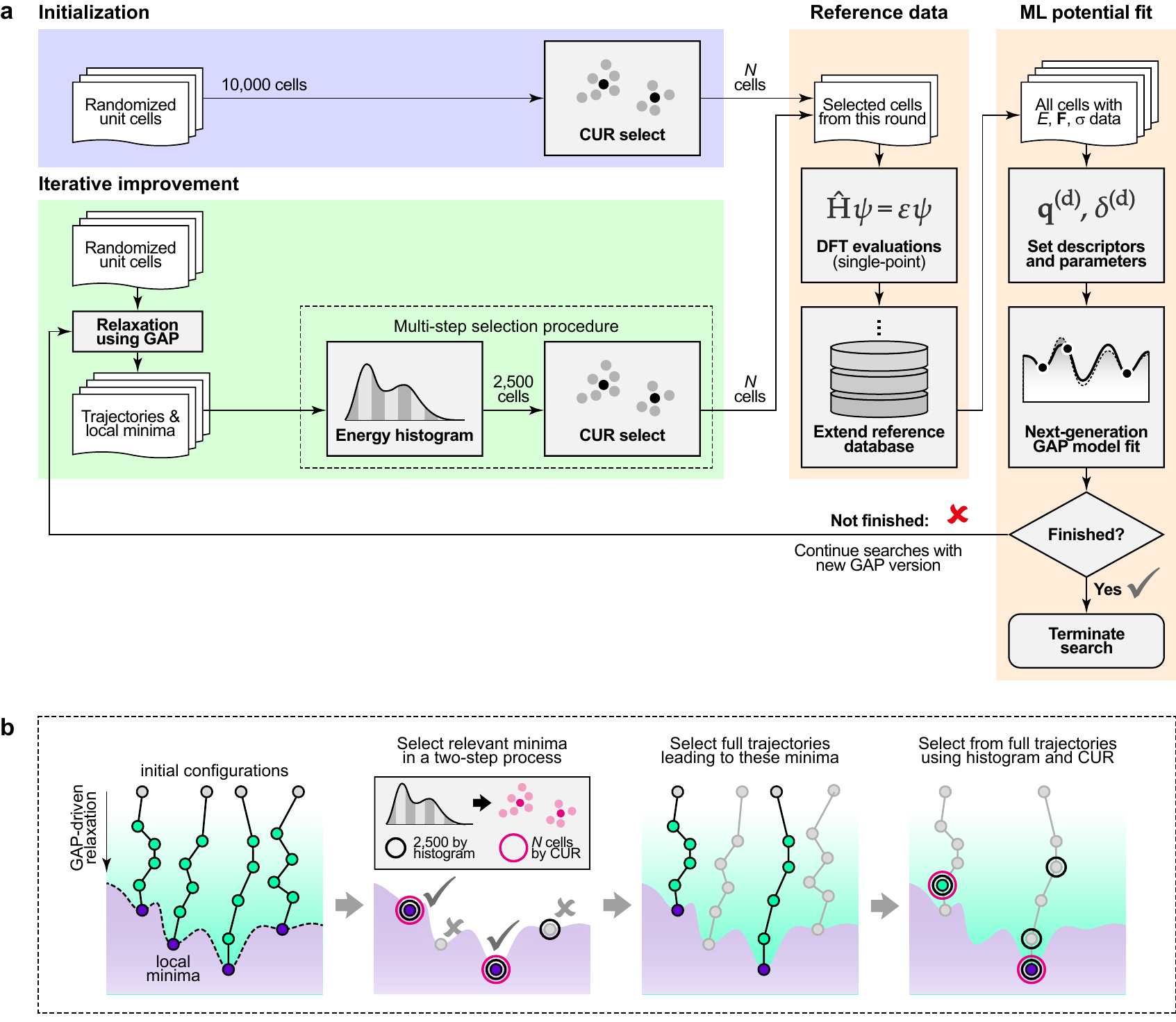} 
    \caption{  
    An automated protocol that iteratively explores structural space and fits machine learning (ML) based interatomic potentials. ({\bf a}) General overview of the approach. From an ensemble of randomized unit cells ({\em blue}), we select the most relevant ones using the leverage-score CUR algorithm. Selected cells are evaluated with single-point DFT computations and used to fit an initial Gaussian Approximation Potential (GAP) ({\em orange}). Then, this potential is used to relax a new ensemble of randomized cells ({\em green}), selecting again the most relevant snapshots, and repeating the cycle. ({\bf b}) Illustration of the multi-step selection procedure. We first consider all trajectories in a given generation, sketched by connected points, and select the most relevant local minima (using an energy criterion, the flattened histogram, and then a structural criterion, the CUR clustering). From the trajectories leading to these minima, we then select the most representative cells; these can be intermediates ({\em green}) or end points ({\em purple}) of relaxations. The structures finally selected ({\em magenta}) are DFT-evaluated and added to the database.}
    \label{fig_flowchart}
\end{figure*}

In the context of ML potential fitting, so-called ``active learning'' schemes which detect extrapolation (indicating when the potential moves away from ``known'' configurations) are currently receiving much attention. 
A query-by-committee active-learning approach was suggested in 2012 by Artrith and Behler \cite{Artrith2012}.
More recently, Shapeev and co-workers employed Moment Tensor Potentials \cite{Shapeev2016} with active learning\cite{Podryabinkin2017} to explore the PES and to fit ML potentials \cite{Podryabinkin2019, Gubaev2019}, and E and co-workers described a generalized active-learning scheme for deep neural network potentials \cite{Zhang2018}. 
So far, these studies mainly focused on specific intermetallic systems, namely, Al--Mg \cite{Zhang2018} and Cu--Pd, Co--Nb--V, and Al--Ni--Ti \cite{Gubaev2019}, respectively.
Furthermore, Podryabinkin et al.\ showed that their approach can identify various existing and hypothetical boron allotropes \cite{Podryabinkin2019}.
Finally, Jinnouchi et al.\ demonstrated how ab initio molecular dynamics (AIMD) simulations of specific systems can be sped up by active learning of the computed forces (in a modified GAP framework), using the predicted error of the Gaussian process to select new datapoints and to improve the speed of AIMD \cite{Junnouchi2019, Junnouchi2019a}.

In this work, we present an efficient and unified approach for fitting ML potentials by GAP-RSS, exploring structural space from the beginning ({\em de novo}) by ML-driven searching and similarity-based distance metrics, all without any prior input of what structures are or are not relevant.
We demonstrate the ability to cover a broad range of structures and chemistries, from graphite sheets to a  densely packed transition metal.
Our work provides conceptual insight into how computers can discover structural chemistry based on data and distance metrics alone, and it paves the way for a more routine application of ML potentials in materials discovery.

\section*{Results}

\subsection*{A unified framework for exploring and fitting structural space}

The overarching aim is to construct a ML potential with minimal effort: both in terms of computational resources and in terms of input required from the user. In regard to the former, we use only single-point DFT computations to generate the fitting database \cite{B_GAP}. In regard to the latter, we define general heuristics wherever possible, such that neither the protocol nor its parameters need to be manually tuned for a specific system. The ML architecture is based on a hierarchical combination of two-, three-, and many-body descriptors \cite{amoC}. The two parameters that need to be set by the user are a ``characteristic'' distance and whether the material is primarily covalent or metallic.  For the distance, we choose tabulated covalent (for C, B, and Si)\cite{Cordero2008} or metallic (for Ti) radii, depending on the nature of the system. These define the volume of the initial structures and the cutoffs for the ML descriptors (Methods section).

\begin{figure}
    \centering
    \includegraphics{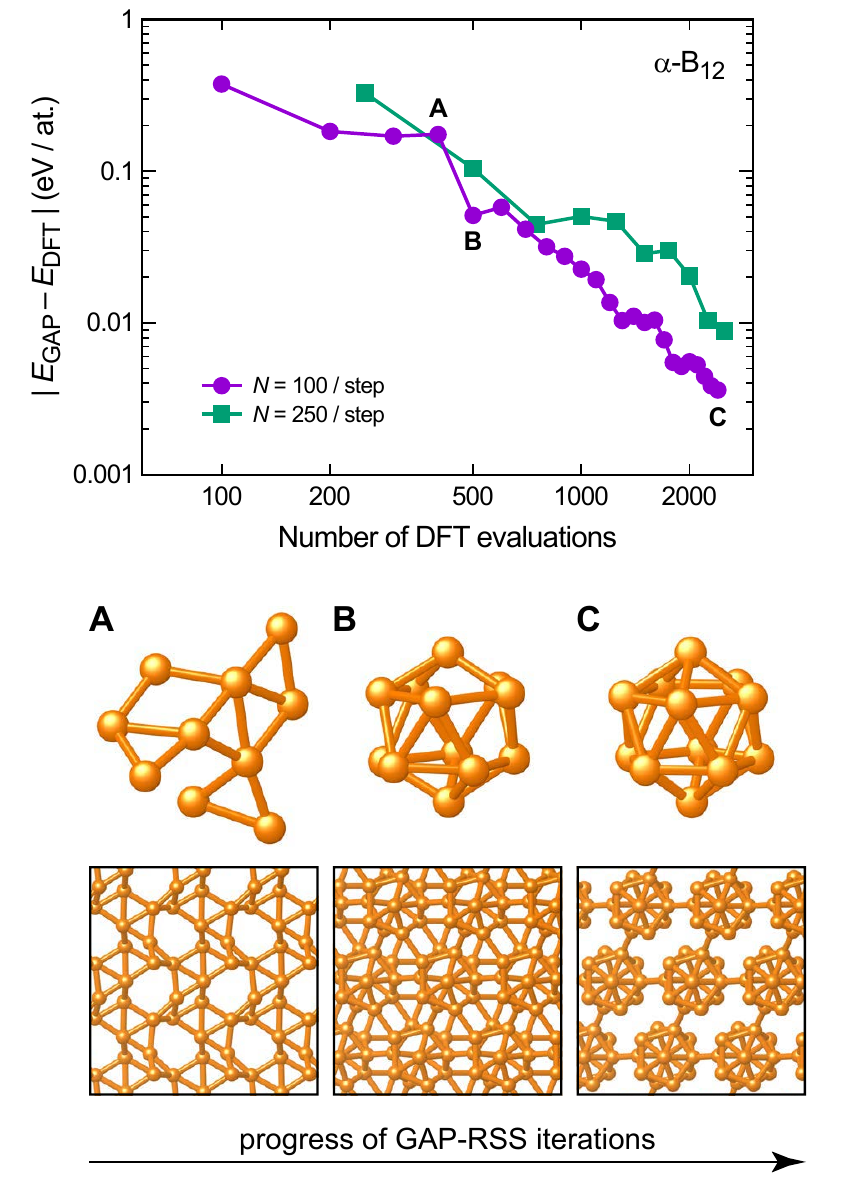} 
    \caption{``Learning'' the crystal structure of $\alpha$-rhombohedral boron.
    {\em Top:} Error of iteratively generated GAP-RSS models, for the energy of the optimised ground-state structure of $\alpha$-B$_{12}$, referenced to DFT. Two independent runs are compared in which $N = 100$ (purple) or $N = 250$ (green) structures per iteration are fed back into the database. {\em Bottom:} Evolution of the B$_{12}$ icosahedron as the defining structural fragment. For three points of the $N = 100$ cycles, having completed 400 (``A''), 500 (``B''), and 2,500 (``C'') DFT evaluations in total, the respective lowest-energy structure (at the DFT level) from this iteration is shown, as visualized using VESTA \cite{Momma2011}. 
    Bonds between atoms are drawn using a cut-off of 1.9 \AA{};
    note that there are further connections between the B$_{12}$ icosahedra with slightly larger B$\cdots$B distances.}
    \label{fig_aB12}
\end{figure}

Our approach is based on an iterative cycle, as shown in the diagram in Fig.~\ref{fig_flowchart}a. We generate ensembles of randomized structures as in the AIRSS framework \cite{Pickard2006, Pickard2011}, a structure-searching approach that is widely used in physics, chemistry, and materials science \cite{Pickard2008, Marques2011, Stratford2017}. In the first iteration, we generate 10,000 initial structures, from which we select the $N$ most diverse ones using the leverage-score CUR algorithm \cite{Mahoney2009}. The distance between candidate structures, therein, is quantified by the Smooth Overlap of Atomic Positions (SOAP) descriptor \cite{Bartok2013}, which has been widely used in GAP fitting \cite{amoC, Bartok2018} and in structural analysis \cite{De2016, Mavracic2018, Caro2018a}. While SOAP is normally used to discriminate between pairs of individual atoms and thus of local configurations, we here use a {\em configuration-averaged} SOAP descriptor that compares entire unit cells to one another (Methods section) \cite{Mavracic2018}. We find this to be crucial for selecting the most representative structures, of which we can only evaluate a small number ($\ll 10,000$) with DFT. We also generate dimer configurations in vacuum at a wide range of bond lengths. 

With the starting configurations in hand, we perform single-point DFT computations and fit an initial (coarse) potential to the resulting data; in subsequent iterations, we extend the database and thereby refine the potential \cite{B_GAP}. In each iteration, we start from the same number of initial structures, and minimize their enthalpy using the GAP from the previous iteration.  We then select the $N$ most relevant and diverse configurations from the full set of configurations seen throughout the minimization trajectories, for which we employ a combination of Boltzmann-probability biased flat histogram sampling (to focus on low-energy structures) and leverage-score CUR (to select the most diverse structures among those), as illustrated in Fig.~\ref{fig_flowchart}b.  These selected configurations are evaluated using single-point DFT calculations and added to the fitting database.

The iterative procedure runs until the results are satisfactory.
Here we terminate our searches after 2,500 DFT data points have been collected, and our results show this to be sufficient to discover and describe all structures discussed in the present work. 
Other quality criteria, such as based on the distribution of energies in the database \cite{B_GAP}, might be defined as well; the generality of our approach is not affected by this choice.

\begin{figure*}
    \centering
    \includegraphics[width=17.2cm]{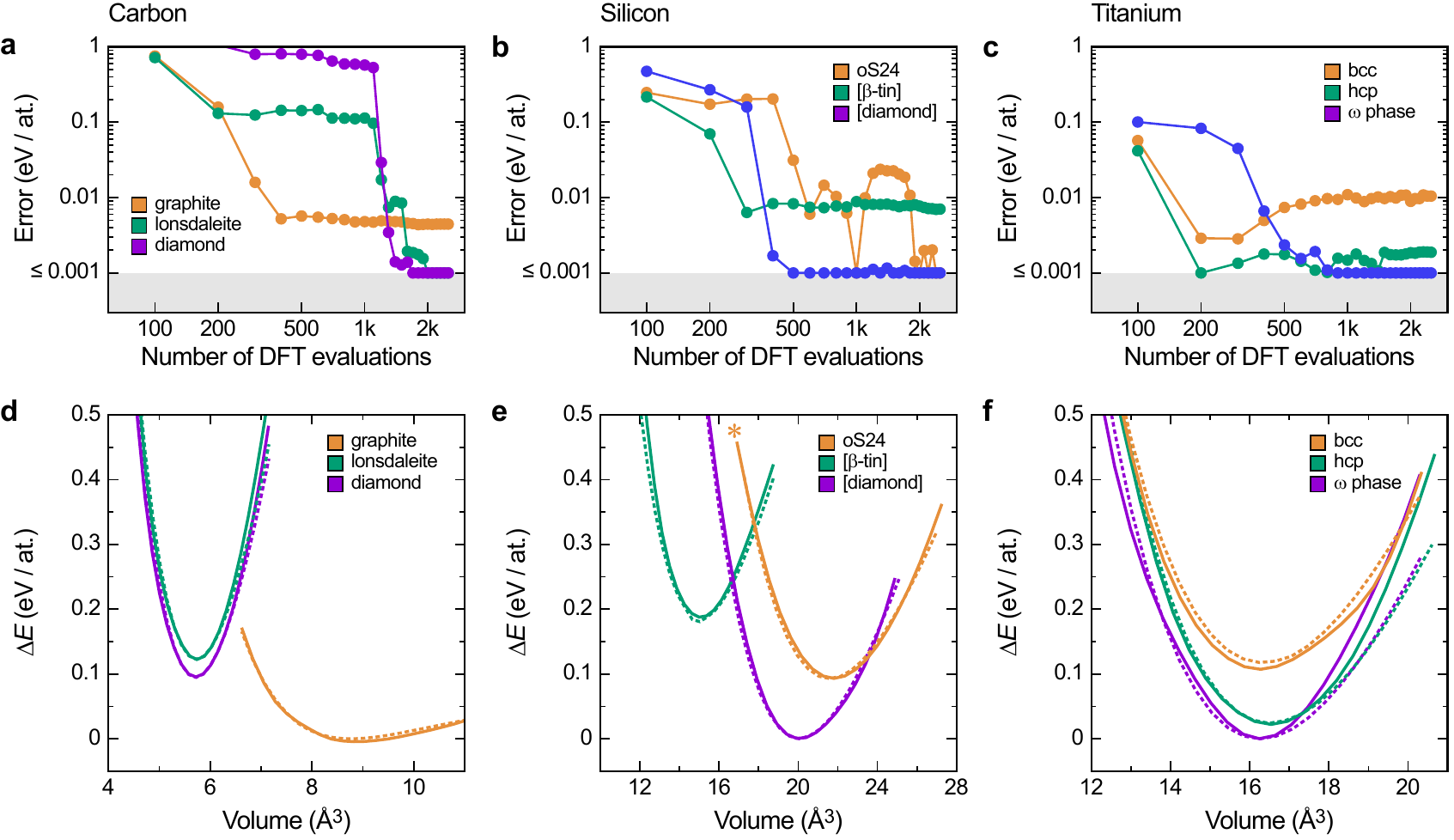}
    \caption{``Learning'' diverse crystal structures without prior knowledge, including textbook examples of an insulator (carbon), a semiconductor (silicon), and a metal (titanium).
    ({\bf a}--{\bf c}) Energy error, defined as the difference between DFT- and GAP-computed energies for structures optimized with the respective method. 
    GAP-RSS models that deviate from the DFT result by less than 1~meV / atom are considered to be fully converged and therefore their errors are drawn as a constant minimum value to ease visualization.
    ({\bf d}--{\bf f}) Energy--volume curves computed with the final GAP-RSS model (solid lines) and the DFT reference method (dashed lines). The open-framework oS24 structure, at high pressure, collapses into a more densely packed phase (``$\ast$''; see SI for details).
    All energies are referenced to the DFT result for the respective most stable crystal structure.}
    \label{fig_EVs}
\end{figure*}

We demonstrate the method for boron, one of the most structurally complex elements \cite{Albert2009}. With the exception of a high-pressure $\alpha$-Ga type phase, all relevant boron allotropes contain  B$_{12}$ icosahedra as the defining structural unit \cite{Albert2009}. Boron has been the topic of structure searches with DFT \cite{Oganov2009, Wu2012, Mannix2015, Ahnert2017} and, more recently, with ML potentials for bulk allotropes \cite{B_GAP, Podryabinkin2019} and gas-phase clusters \cite{Tong2018}. 
Our previous work showed how the PES for boron can be fitted in a ML framework \cite{B_GAP}, leading to the first interatomic potential able to describe the different allotropes. However, at that time, we generated and fed back 250 cells per iteration (without further selection), and added the structure of $\alpha$-B$_{12}$ manually at a later stage.\cite{B_GAP} 

Our new protocol ``discovers'' the structure of $\alpha$-B$_{12}$ in a self-guided way, as illustrated in Fig.\ \ref{fig_aB12}. The increasingly accurate description of the B$_{12}$ icosahedron is reflected in a gradually lowered energy error, falling below the 10~meV/atom threshold with less than 2,000 DFT evaluations, and below 4~meV/atom once the cycle is completed. This improvement is best understood by inspecting the respective lowest-energy structures that enter the database in a given iteration (Fig.\ \ref{fig_aB12}). The lowest-energy structure at point {\bf A} already contains several three-membered rings, but no B$_{12}$ icosahedra yet. With one more iteration, there is a sharp drop in the GAP error (from 175 to 51~meV/at.), concomitant with the first appearance of a rather distorted $\alpha$-B$_{12}$ structure ({\bf B}). The final database has seen several instances of the correctly ordered structure ({\bf C}). 

Our experiment suggests that adding $N=100$ structures at each step slightly outperforms a similar cycle with $N=250$, although both settings lead to satisfactory results. In the remainder of this paper, we will perform all GAP-RSS searches with $N=100$ and up to a total of 2,500 single-point DFT evaluations.

\subsection*{Learning diverse crystal structures}

Our method is not restricted to a particular chemical system. To demonstrate this, we now apply it to three prototypical materials side by side: carbon, silicon, and titanium, which all exhibit multiple crystal structures.

In carbon (Fig.~\ref{fig_EVs}a), both the layered structure of graphite and the tetrahedral network of diamond are correctly ``learned'' during our iterations. For graphite, the energy error reaches a plateau after only a few hundred DFT evaluations; for diamond, the initial error is very large, and after a dozen or so iterations we observe a rapid drop---concomitant with a drop in the error for the structurally very similar lonsdaleite (``hexagonal diamond''). The final prediction error is well below 1~meV/atom for the $sp^{3}$ bonded allotropes, and on the order of 4~meV/atom for graphite. We have previously shown that the forces in diamond show higher locality than those in graphite, making their description by a finite-ranged ML potential easier \cite{amoC}, given that sufficient training data are available. We also note that our method captures the difference between diamond and lonsdaleite very well: its value is 27~meV/atom with the final GAP-RSS version, and 28~meV/atom with DFT. 

In silicon (Fig.~\ref{fig_EVs}b), the ground-state (diamond-type) structure is very quickly learned, more quickly so than diamond carbon, which we ascribe to the absence of a competing threefold-coordinated phase in the case of Si. We further test our evolving potentials on the high-pressure form, the $\beta$-tin type allotrope (space group $I4_{1}/amd$), which is easily discovered; the larger residual error for $\beta$-Sn-type than for diamond-type Si is consistent with previous studies using a manually tuned potential \cite{Bartok2018}. We also test our method on a recently synthesized open-framework structure with 24 atoms in the unit cell (oS24) \cite{Kim2015}, which consists of distorted tetrahedral building units that are linked in different ways, which the potential has not ``seen''. Still, a good description is achieved after a few iterations. 

In titanium (Fig.~\ref{fig_EVs}c), a hexagonal close packed (hcp) structure is observed at ambient conditions; however, the zero-Kelvin ground state has been under debate: depending on the DFT method, either hcp or the so-called $\omega$ phase is obtained as the minimum. Our method clearly reproduces the qualitative and quantitative difference between the two allotropes (22~meV/atom with the final GAP-RSS iteration versus 24~meV/atom with DFT) at the computational level we use.

Looking beyond the minimum structures, the DFT energy--volume curves are, by and large, well reproduced by GAP-RSS; see Fig.~\ref{fig_EVs}d--f. There is some deviation at large volumes for hcp and $\omega$-type Ti, but this is an acceptable issue as these regions of the PES are not as relevant, corresponding to negative external pressure. If one were interested in very accurate elastic properties, one would choose to include less dense structures by modifying the pressure parameters (Methods section, Eq.\ \ref{eq:pressure}). Indeed, it was recently shown that a ML potential for Ti, fitted to a database of 2,700 structures built from the phases on which we test here ($\omega$, hcp, bcc) and other relevant structures can make an accurate prediction of energetic and elastic properties \cite{Takahashi2017}. 

\begin{figure}
    \centering
    \includegraphics{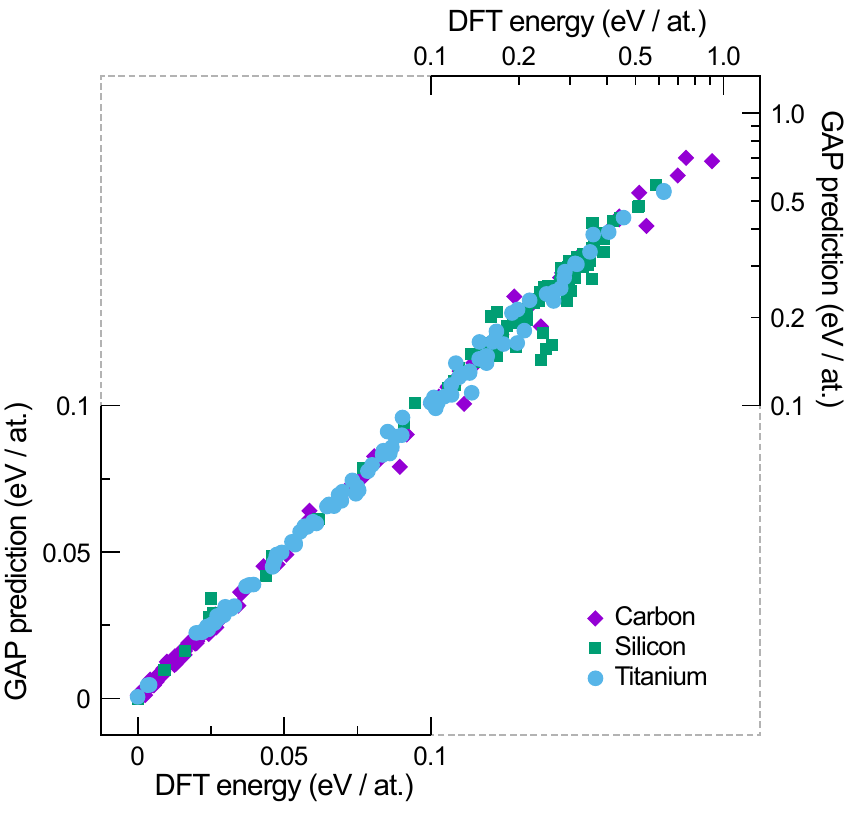}
    \caption{Scatter plots of predicted versus DFT energies for ensembles of structures added to the reference databases in the final iteration. Note that the energy scale is continuous, but it changes from linear to logarithmic scaling at 0.1 eV/at., allowing us to visualize both low- and higher-energy regions.}
    \label{fig_errors_square}
\end{figure}

\subsection*{Entire potential-energy landscapes}

While the most relevant crystal structures for materials are usually well known and available from databases, we show that our chemically ``agnostic'' approach is more general. In Fig.\ \ref{fig_errors_square}, we show an energy--energy scatter plot for the last set of GAP-RSS minimizations, evaluated with DFT and with the preceding GAP version, and again across three different chemical systems. 
We survey both the low- and higher-energy regions of the PES---up to 1 eV per atom, which is very roughly the upper stability limit at which crystalline carbon phases may be expected to exist \cite{Aykol2018}.

\begin{figure*}
    \centering
    \includegraphics[width=16.8cm]{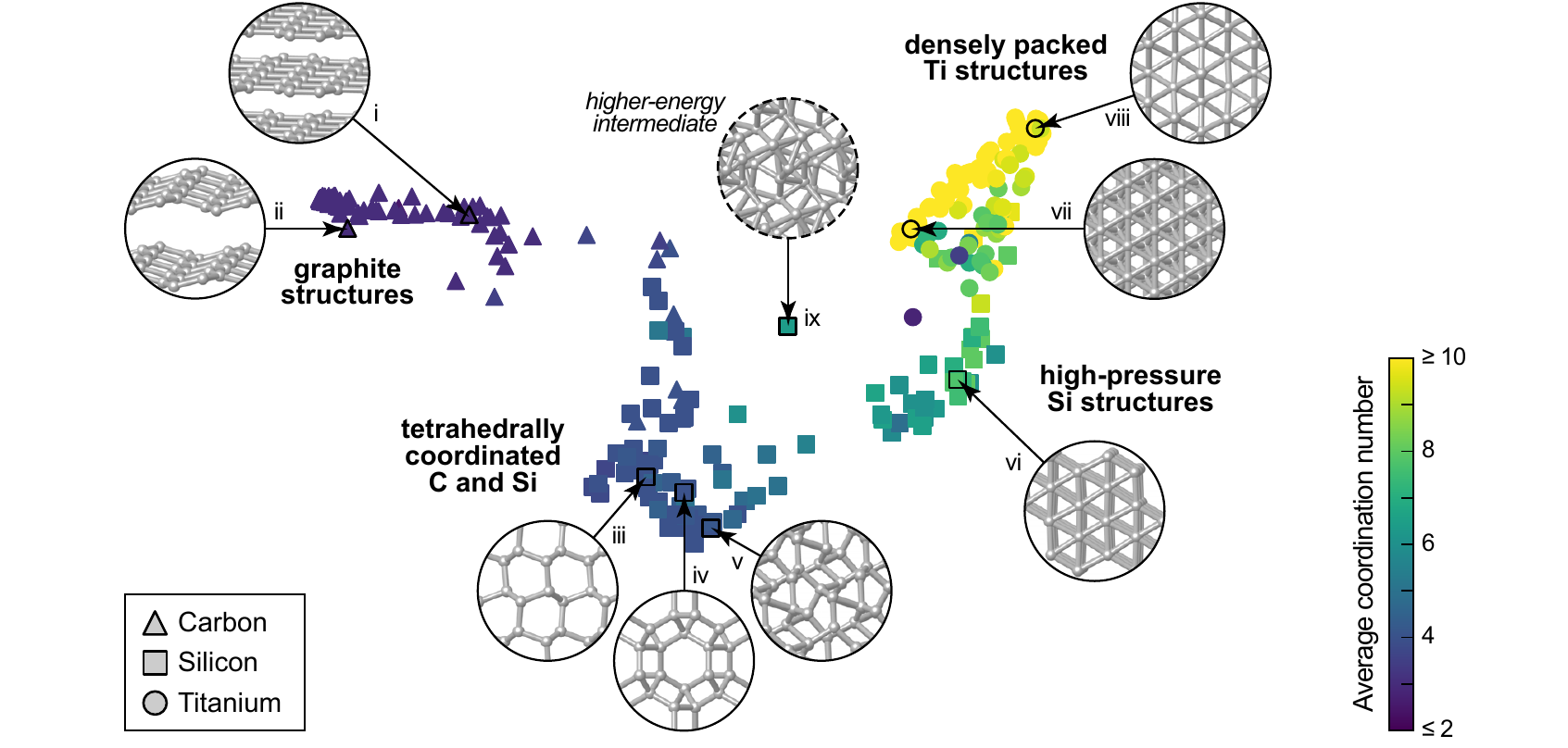}
    \caption{ Visualizing the highly diverse structures, both at low and relatively high energies above the global minimum, that have been explored by GAP-RSS and added to the reference database in the last iteration. 
     A similarity map compares three systems side-by-side (carbon, {\em triangles}; silicon, {\em squares}; titanium, {\em circles}), as described in the text. The resulting plot (with arbitrary axis values) emphasizes relationships between the different databases. The structures, ``learned'' from scratch by our protocol, range all the way from threefold-coordinated graphite, fourfold-coordinated (sp$^{3}$-like) allotropes of C and Si, onward to high-pressure Si structures and finally densely packed variants of Ti. A higher-energy structure ($\approx 0.6$ eV/at.\ above diamond-type silicon) from an earlier step in a minimization trajectory is included as an example, as enclosed by a dashed line.}
    \label{fig_mds}
\end{figure*}

To analyze and understand the outcome of these searches in structural and chemical terms, we compute the distance between any two structures A and B as
\begin{align}
    d_{\rm AB}=\sqrt{2-2k_{\rm AB}},
\end{align}
where $k_{\rm AB}$ is again the configuration-averaged SOAP kernel for A and B,i.e.\ a distance between two entire unit cells (Methods section). We then use a dimensionality reduction technique to draw a two-dimensional structural map reflecting these distances. Such SOAP-based maps have been used with success to analyze structural and chemical relationships in different materials datasets \cite{De2016, Engel2018, Caro2018a}. Here, we use them to illustrate how different materials (including their allotropes as known from chemistry textbooks) are related in structural space. 

To compare different materials with inherently different {\em absolute} bond lengths, we re-scale their unit cells such that the minimum bond length in each is $r_{0} = 1.0$~\AA{}, inspired by approaches for topological analyses of different structures \cite{DelgadoFriedrichs2003}. We then use SOAP ($r_{\rm cut}=2.5~r_{0}$, $\sigma_{\rm at}=0.1~r_{0}$) to determine the distance of all pairs of cells in this set, and use principal component analysis to represent this dataset in a 2D plane. Figure \ref{fig_mds} shows the resulting plot, in which we have encoded the species by symbols and the average coordination number by color. (Coordination numbers are determined by counting nearest neighbors up to $1.2~r_{0}$.)

The results fall within four groups, moving from the left to the right through Fig.\ \ref{fig_mds}. The first group is given by graphite-like structures; they are three-fold coordinated and only carbon structures (circles) are found there. Roman numerals in Fig.\ \ref{fig_mds} indicate examples, and in this first group we observe flat (i) and buckled (ii) graphite sheets. In the second group, we have four-fold coordinated (``diamond-like'') networks, made up by both carbon and silicon (recall that we are using a normalized bond length, so diamond carbon and diamond-type silicon will fall on the same position in the plot). The structures that are shown as insets are characteristic examples; from left to right, there is a distorted lonsdaleite-type structure (iii), the well-known {\bf unj} framework (also referred to as the ``chiral framework structure'' in group-14 elements \cite{Pickard2010}; iv), and a more complex $sp^{3}$-bonded allotrope (v). While the axis values in our plot are arbitrary, they naturally reflect the structural evolution toward higher coordination numbers, and therefore we next observe a set of high-pressure silicon structures (squares), such as the simple-hexagonal one (vi), with an additional contribution from lower-coordinated titanium structures (circles). Finally, there is a set of densely packed structures, all clustered closely together; these are titanium structures including hcp (vii) and the $\omega$ type (viii). In the center of the plot, there is a structure that bears resemblance to none of the previously mentioned ones (ix), an energetically high-lying and strongly disordered intermediate from a relaxation trajectory that was added to the reference database, rather than a local minimum. This dissimilarity is reflected in relatively large distances from other entries in the SOAP-based similarity map.

\section*{Discussion}

We have shown that automated protocols can be designed for generating structural databases and fitting potential-energy surfaces of materials in a self-guided way. 
This allows for the generation of ML-based interatomic potentials with minimal effort, both in terms of computational and user time, and it represents a step toward wide applicability of these techniques in computational materials science.
Once a core (RSS-based) database has been constructed, it can be readily improved by adding defect, surface, and liquid/amorphous structural models, while at the same time being sufficiently robust to avoid unphysical behavior---even when taken to the more extreme regions of configuration space that are explored early on during RSS. 

We targeted here the space of three-dimensional inorganic crystal structures, but conceptually similar approaches may be useful for nanoparticles \cite{Artrith2014, Kolsbjerg2018}. Finally, organic (molecular) materials are also beginning to be described very reliably with ML potentials \cite{Smith2017, Chmiela2018}, and an interesting open question is how to use the structural diversity inherent in RSS in the context of organic solids \cite{Zilka2017}.

\section*{Methods}

\small

\subsection*{Interatomic potential fitting}

To fit interatomic potentials, we use the Gaussian Approximation Potential (GAP) ML framework \cite{Bartok2010} and the associated computer code, which is freely available for non-commercial research at http://www.libatoms.org. Compared to previous work, we have here developed suitable heuristics to automate and generalize the choice of fitting parameters where possible.

We use a linear combination of 2-, 3-, and many-body terms following Refs.\ \citenum{Bartok2015} and \citenum{amoC}, with defining parameters given in Table \ref{tab:GAP_hypers}. 
The 2-body (``2b'') and 3b descriptors are scalar distances and symmetrized three-component vectors, respectively. For the many-body term, we use the Smooth Overlap of Atomic Positions (SOAP) kernel \cite{Bartok2013}, which has been used to fit GAPs for diverse systems \cite{Szlachta2014, amoC, Bartok2018, Mocanu2018}. The overall energy scale of each descriptor's contribution to the predicted energy (controlled by the parameter $\delta$) \cite{Bartok2015} is set automatically in our  protocol. The 2b value is set from the variance of energies in the fitting database, the 3b value is set from the energy error between a 2b only fit and the fitting database, and the SOAP value is set from the energy error for a 2b+3b only fit. 

The cutoffs for the three types of descriptors are expressed in terms of the characteristic radius $r$ (Table \ref{tab:GAP_hypers}): that for 2b is longest range, while that for 3b is shortest (intended to capture only nearest neighbors), and the SOAP is intermediate in range. The resulting cutoff settings are listed in Table~\ref{tab:GAP_hypers}, the characteristic radii $r$ for the systems studied here being 0.84, 0.76, 1.11, and 1.47~{\AA} for B, C, Si, and Ti, respectively. 

The weights on the energies, forces, and stresses that are fit are set by diagonal noise terms in Gaussian process regression \cite{Bartok2010}. We set these according to the reference energy of a given structure, to make the fit more accurate for relatively low-energy structures at each volume while providing flexibility for the higher-energy regions. The values are piecewise-linear functions in $\Delta E$, which is the per-atom reference energy difference relative to the same volume on the convex hull bounding the set of $(V,E)$ points from below (in energy).  For the energy the error $\sigma_E$ is 1~meV/atom for $\Delta E \le 0.1$~eV, 100~meV/atom for $\Delta E \ge 1$~eV, and linearly interpolated in between.  For forces the corresponding $\sigma_F$ values are 31.6 and 316~meV/{\AA}, and for virials the $\sigma_V$ values are 63.2 and 632~meV/atom.

\subsection*{Comparing structures}

We also use SOAP, although with different parameters ($n_{\rm max} = l_{\rm max} = 12$, $\sigma_{\rm at}=0.0875$~\AA{}, $r_\mathrm{cut}=10.5$~\AA), to compare the similarity of environments (as proposed in Ref.~\citenum{De2016}) in selecting from which data to train (in the CUR step). We obtain what we call a ``configuration-averaged'' SOAP by averaging over all atoms in the cell.
In the SOAP framework \cite{Bartok2013}, the neighbor density of a given atom $i$ is expanded using a local basis set of radial basis functions $g_{n}$ and spherical harmonics $Y_{lm}$,
\begin{align}
  \label{eq:soap_rho_expansion}\nonumber 
  \rho_{i}({\vec{r}}) 
   =& \sum_j \exp \left(-|r-r_{ij}|^2/2\sigma_\textrm{at}^2\right) \\
   =& \sum_{nlm} c^{(i)}_{nlm} \, g_{n}(r) Y_{lm}(\hat{\vec{r}}),
\end{align}
where $j$ runs over the neighbours of atom $i$ within the specified cutoff (including $i$ itself). 
To obtain a similarity measure between unit cells, rather than individual atoms, we then average the expansion coefficients over all atoms $a$ in the unit cell,
\begin{equation}
\bar{c}_{nlm} = \frac{1}{N} \sqrt{\frac{8 \pi^{2}}{2l+1}} \sum_{i}  c^{(i)}_{nlm},  
\end{equation}
and construct the rotationally invariant power spectrum for the entire unit cell \cite{Mavracic2018},
\begin{equation}
 \bar{p}_{nn'l} =  \sum_{m}  \left( \bar{c}_{nlm} \right)^{\ast} \bar{c}_{n'lm}.
\end{equation}
Note that this is not equal to the average of the usual atomic SOAP power spectra 
used to describe the atomic neighbor environments. 
The final kernel to compare two cells, A and B, is then
\begin{equation}
\label{eq:dot_prod_kernel}
    k_{\rm AB} = \left( \sum_{nn'l} \bar{p}^{\rm \,(cell\; A)}_{nn'l} \bar{p}^{\rm \,(cell\; B)}_{nn'l} \right)^{\zeta},
\end{equation}
where $\zeta$ is a small integer number (here, $\zeta=4$).

\begin{table}
\caption{Hyperparameters for descriptors that we use in GAP fitting. For all descriptors: Gaussian width $\sigma_{\rm at}$ (squared-exponential kernel for 2- and 3-body; atomic density width for SOAP); number of sparse points $N_\mathrm{sp}$. For SOAP only: number of radial functions $n_\mathrm{max}$ and angular momenta $l_\mathrm{max}$, and kernel exponent $\zeta$.  Cutoffs $r_\mathrm{cut}$ are expressed in terms of the characteristic radius $r$, listed for each material in the Interatomic potential fitting subsection.}
\label{tab:GAP_hypers}
\begin{tabular}{lccccccc}
\hline 
\hline 
 & & & & & & \multicolumn{2}{c}{$r_{\rm cut}$ (\AA{})} \\
 &  $\sigma_\textrm{at}$ (\AA) & $N_\mathrm{sp}$ & $n_\mathrm{max}$ & $l_\mathrm{max}$ & $\zeta$ &  {\footnotesize (covalent)} & {\footnotesize (metallic)} \\ 
\hline 
 2-body & 0.5 & 30 & & & & $9.0 \, r$ & $8.2 \, r$ \\ 
 3-body & 1.0 & 100 & & & & $2.925 \, r$ & $2.665 \, r$ \\ 
\hline 
 SOAP & 0.75 & 2000 & 8 & 8 & 4 & $ 4.5 \, r$ & $4.1 \, r$ \\ 
\hline 
\hline 
\end{tabular}
\end{table}

\subsection*{Iterative generation of reference data}

Randomized atomic positions are generated using the {\tt buildcell} code of the AIRSS package version 0.9, available at https://www.mtg.msm.cam.ac.uk/Codes/AIRSS. The positions are repeated by 1--8 symmetry operations, and the cells contain 6--24 atoms. A minimum separation is also set, with a value of $1.8 r$. The volumes per atom of the random cells are centered on $V_{0} = 14.5 \, r^{3}$ for covalent, and $V_{0} = 5.5 \, r^{3}$ for metallic systems. In the initial iteration, half of the structures are generated from the {\tt buildcell}-default narrow range of volumes, and half from a wider range, $\pm 25$\% from the heuristic value.  In all later iterations, only the default narrow range is used.  The wide volume range configurations are meant to simply span a wide range of structures \cite{B_GAP}, and use only even numbers of atoms.  The narrow volume range configurations are meant to be good initial conditions for RSS, and so for 80\% (20\%) of the seed structures, we choose even (odd) numbers of atoms, respectively. This is because for most known structures, the number of atoms in the conventional unit cell is even (eight for diamond and rocksalt, for example), although for some it is odd, including the $\omega$ phase \cite{Sikka1982}. Biasing initial seeds toward distributions that occur in nature is a central idea within the AIRSS formalism \cite{Pickard2011}.    

With the initial potential available, we then run structural optimizations by relaxing the candidate configurations with a preconditioned LBFGS algorithm~\cite{Packwood2016} to minimize the enthalpy until residual forces fall below 0.01~eV/\AA. As in Ref.~\citenum{GAP-RSS_Faraday}, we employ a random external pressure $p$ with probability density
\begin{equation}
    \label{eq:pressure}
    P(p/p_0) = \frac{1}{\beta} \exp \left( - \frac{1}{\beta} \, p/p_0 \right) \, ,
\end{equation}
here with $p_{0}=1$~GPa, and $\beta=0.2$. This protocol ensures that there is a small but finite external pressure, and also some smaller-volume structures are included in the fit \cite{B_GAP, GAP-RSS_Faraday}. We choose the same pressure range for all materials, for simplicity, although this value could be adjusted depending on the pressure region of interest.

The selection of configurations for DFT evaluation and fitting at each iteration involves a Boltzmann-biased flat histogram and leverage-score CUR, as illustrated in Fig.~\ref{fig_flowchart}. To compute the selection probabilities for the flat histogram stage, the distribution of enthalpies (each computed using the pressure at which the corresponding RSS minimization was done) is approximated by the numpy~\cite{numpy} histogram function, with default parameters.  The probability of selecting each configuration is inversely proportional to the density of the corresponding histogram bin, multiplied by a Boltzmann biasing factor.  The biasing factor is exponential in the energy per atom relative to the lowest energy configuration, divided by a temperature of 0.3~eV for the first iteration, 0.2~eV for the second, and 0.1~eV for all remaining iterations. The leverage-score CUR selection is based on the singular-value decomposition of the square kernel matrix using the SOAP descriptors (with the dot-product kernel and exponentiation by $\zeta$, Eq.~\ref{eq:dot_prod_kernel}).  Applying the same algorithm to the rectangular matrix of SOAP descriptor vectors was significantly less effective.

\subsection*{Computational details}

Reference energies and forces were obtained using DFT, with projector augmented-waves (PAW)~\cite{Blochl1994,Kresse1999} as implemented in the Vienna Ab Initio Simulation Package (VASP) \cite{Kresse1996}. Valence electrons were described by plane-wave basis sets with cutoff energies of 500 (B), 800 (C), 400 (Si), and 285 eV (Ti), respectively. Exchange and correlation were treated using the PBEsol functional \cite{Perdew2008} for all materials except carbon, where the opt-B88-vdW functional \cite{Dion2004,RomanPerez2009,Klimes2011}  was chosen to properly account for the van der Waals interactions in graphitic structures. Benchmark data for energy--volume curves were obtained by scaling selected unit cells within given volume increments and optimizing while constraining the volume and symmetry of the cell.

\normalsize

\section*{Data access statement}
Data supporting this publication will be made openly available via https://www.repository.cam.ac.uk.\\

\section*{Author contributions}
N.B., G.C., and V.L.D. jointly designed the research, developed the approach, and analyzed the data. N.B. developed the computational framework and performed the computations with input from all authors. V.L.D. wrote the paper with input from all authors.

\begin{acknowledgements}
We thank Profs.~C.~J.~Pickard and D.~M.~Proserpio for ongoing valuable discussions.
N.B. acknowledges support from the Office of Naval Research through the U. S. Naval Research Laboratory's core basic research program.
G.C. acknowledges EPSRC grant EP/P022596/1.
V.L.D. acknowledges a Leverhulme Early Career Fellowship and support from the Isaac Newton Trust.
\end{acknowledgements}

\cleardoublepage

\onecolumngrid
\appendix 

\renewcommand\thefigure{S\arabic{figure}}    
\setcounter{figure}{0}    

\renewcommand\thetable{S\arabic{table}}    
\setcounter{table}{0}    

\renewcommand\thepage{S\arabic{page}}    
\setcounter{page}{1}    

\begin{center}
    \large {\bf Supplementary Information for
    ``De novo exploration and self-guided learning of potential-energy surfaces''}\\[8mm]
\end{center}

\twocolumngrid

\begin{figure}[H]
    \centering
    \includegraphics{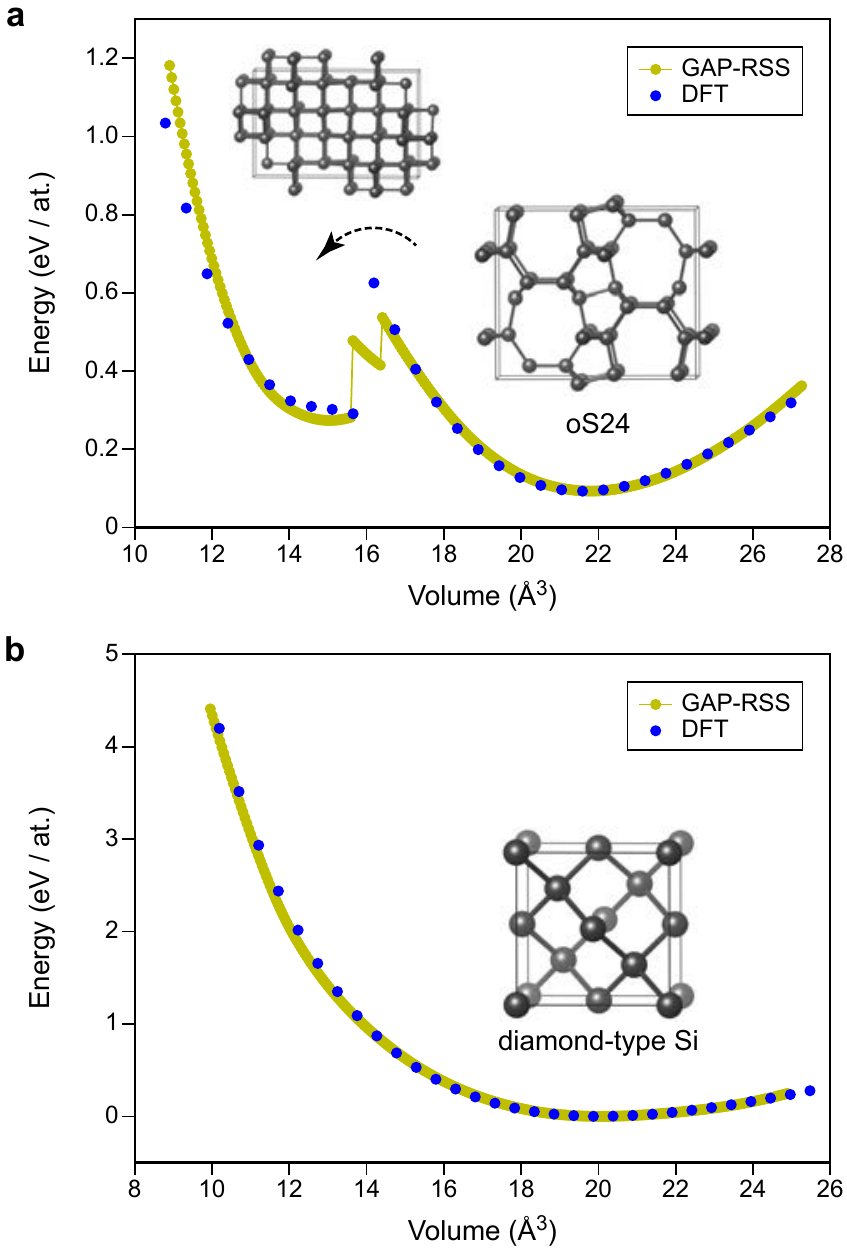} 
    \caption{Silicon allotropes at high pressure ({\em as a supplement to Fig.\ \ref{fig_EVs}e}). We here show energy--volume curves as in the main text, but with
    smaller increments and extending down to much smaller unit-cell volumes (that is,
    higher-pressure regions), down to 50\% of the equilibrium volume. In the case of the open-framework silicon structure oS24
    (Ref.\ \citenum{Kim2015}; panel {\bf a}), a collapse is observed at $\approx 16$ \AA{}$^{3}$, both in the
    DFT reference computation and in the GAP-RSS prediction.
    We also performed a test for diamond-type silicon, compressing it to similarly small volumes (and much more strongly than shown in Fig.\ \ref{fig_EVs}e). Over the full range of volumes studied, good agreement is observed between the DFT reference and the GAP-RSS prediction (panel {\bf b}), which indicates a robust ``learning'' of repulsion at very small interatomic distances.}
    \label{fig_SI_silicon}
\end{figure}

\begin{table}[H]
\caption{Additional information regarding the selected structures shown as insets in Fig.\ \ref{fig_mds}: external pressure applied for this relaxation trajectory, $p_{\rm ext}$; energy relative to the respective ground state, $\Delta E$; maximum DFT-computed force component $F_{i}$ in this structure.}
\label{tab:SI_structures1}
\small 
\begin{center}
\begin{tabular}{llccc}
\hline 
\hline 
           &   & $p_{\rm ext}$ & $\Delta E$ & max $\left\{F_{i}\right\}$\\
           &   & (GPa) & (eV\,/\,at.) & (eV\,/\,\AA{}) \\
\hline 
(i)    & C (graphite)  & 0.111 & +0.15 & 0.302 \\ 
(ii)   & C (buckled)   & 0.088 & +0.23 & 1.678 \\ 
\hline 
(iii)  & Si (dist. {\bf lon})  & 0.048 & +0.32 & 0.742 \\ 
(iv)   & Si ({\bf unj})        & 0.000 & +0.06 & 0.022 \\ 
(v)    & Si ($sp^{3}$ network) & 0.124 & +0.29 & 0.962 \\ 
(vi)   & Si (simple hex.)      & 0.208 & +0.25 & 0.653 \\ 
\hline 
(vii)  & Ti ($\omega$ phase)   & 0.185 & +0.03 & 0.000 \\ 
(viii) & Ti (hcp)              & 0.324 & +0.05 & 0.238 \\ 
\hline 
(ix)   & Si (RSS intermed.)    & 0.172 & +0.59 & 0.857 \\ 
\hline 
\hline
\end{tabular}
\end{center}
\normalsize 
\end{table}

\begin{table}[H]
\caption{Lattice parameters and atomic coordinates for silicon structure (v), as added to the GAP-RSS reference database in the final iteration (Fig.\ \ref{fig_mds}).}
\label{tab:SI_structures2}
\begin{center}
\begin{tabular}{lrrr}
\hline 
\hline 
 & \multicolumn{3}{c}{$a = b = c =  5.5869$ \AA{}}  \\
 & \multicolumn{3}{c}{$\alpha = \beta = \gamma = 109.08^{\circ}$} \\
\hline 
Si &  -1.68605982 &  -0.41624606 &   3.95802416 \\
Si &   0.15435366 &   4.30188720 &   0.38981490 \\
Si &   4.01815977 &  -0.40190360 &   1.54106451 \\
Si &  -2.28421007 &   2.67006652 &   2.34218566 \\
Si &   3.26751299 &   2.67507753 &  -0.00486561 \\
Si &   1.46783885 &  -1.91088147 &   3.46795123 \\
Si &   1.26796258 &   1.77652573 &   3.00303591 \\
Si &   0.48155278 &   0.67469729 &   1.14050708 \\
\hline 
\hline
\end{tabular}
\end{center}
\end{table}

\begin{table}[H]
\caption{As Table \ref{tab:SI_structures2} but for silicon structure (ix).}
\label{tab:SI_structures3}
\begin{center}
\begin{tabular}{lrrr}
\hline 
\hline 
 & \multicolumn{3}{c}{$a = b = c =  6.5595$ \AA{}}  \\
 & $\alpha = 84.76^{\circ}$ & $\beta = 130.49^{\circ}$ & $\gamma = 116.23^{\circ}$ \\
\hline 
Si &  -1.20744491 &   4.11837921 &   3.06869747 \\
Si &   1.15061078 &   2.70140772 &   0.21706451 \\
Si &  -2.74075733 &  -1.74587757 &   4.91801644 \\
Si &  -2.97936509 &   2.43953892 &   3.67216632 \\
Si &  -1.80907975 &   3.75555505 &   5.55352463 \\
Si &  -2.12808832 &  -0.09478641 &   3.03665790 \\
Si &   1.86693516 &   0.25171580 &   3.77895234 \\
Si &  -0.51394068 &   1.68239975 &   3.54412519 \\
Si &   3.12840615 &   4.18317367 &   0.57109224 \\
Si &   2.03701947 &   0.79695955 &   1.44756468 \\
\hline 
\hline
\end{tabular}
\end{center}
\end{table}

\end{document}